\documentstyle[12pt,psfig]{article}
\def\reference{\parskip 0pt\par\noindent\hangindent 0.5 truecm}

\begin{document}
%
% Title
% Capitalise the title normally - do not use ALL CAPS.
%
\title{Proper Motions in Compact Symmetric Objects}
%

% Authors
% Here comes the author(s) of the paper. Please add the appropriate author
% names for your paper and indicate within the $^...$ the number(s)
% which corresponds to the institute(s) of each author. In this example
% the second author has two institutional affiliations.
% Add or remove authors as required.
% **** IMPORTANT: Leave the closing curly bracket line as is. ******

\author{A.G. Polatidis, $^{1}$ 
 J. E. Conway $^{2}$ 
% THIRD AUTHOR $^{1}$ \and
% FOURTH AUTHOR $^{3}$
} % IMPORTANT: leave this curly bracket as the first character of this line.

% Date - leave this blank.
\date{}
\maketitle

% Institutions
% Here fill in your institute name(s) and address(es)
% The number in $^...$ indicates the author number.  For example
{\center
$^1$ Max Planck Institut f\"ur Radioastronomie, Postfach 2024, D-53010, Bonn, Germany\\[3mm]
$^2$ Onsala Space Observatory, Chalmers Technical University,
         Onsala S-43992, Sweden \\[3mm]
%$^3$ INSTITUTE, STREET, CITY, COUNTRY, CODE\\EMAIL ADDRESS OF AUTHOR\\[3mm]
}

% Abstract
% Simply place your abstract between the \begin{abstract} and
% \end{abstract} commands.
%
\begin{abstract}
We discuss recent measurements of proper
motions of the hotspots of Compact Symmetric Objects. Source expansion 
has been detected in ten CSOs so far and all these objects are very 
young ($\leq 3\times 10^{3}$ years). In a few sources ages have also 
been estimated  from energy supply and spectral ageing arguments 
and these estimates are comparable. This argues that these sources 
are close to equipartition and that standard spectral ageing models apply. 
%The fact that double structures with 
%narrow jets are formed in some sources only a few hundred 
%years after the start of activity is a remarkable result and must 
%strongly constrain the scales and mechanisms by which jets are 
%formed. Ram pressure arguments suggest external medium 
%densities of order 1 cm$^{-3}$ on scales of hundreds of 
%parsecs.  
Proper motions studies are now constraining hotspot
accelerations, side-to-side motions and differences in hotspot 
advance speeds between the two hotspots within sources. 
Although most CSOs are young sources their evolution 
is unclear. There is increasing  evidence that in some objects 
the CSO structure represents a   new phase of activity  
within a recurrent  source. 

\end{abstract}

{\bf Keywords:  Galaxies: active-compact-evolution}
% Place keywords here. Please write all keywords in lower case. PASA uses the
 %standard list of subject 
% headings adopted by The Astrophysical Journal and available from URL:
%   http://www.journals.uchicago.edu/ApJ/keywords_text.html

% A formatting command to add space between the author list and the body
% of the paper when printed. This spacing may be changed as desired.
%\smallskip

%
% Body of paper
%

\section{Introduction}

The term ``Compact Symmetric Objects'' (CSOs) was first coined by
Wilkinson et al (1994) to describe sub-kiloparsec scale sources having
symmetric radio structure. Often these sources are doubles or triples
(Conway et al 1992), where the central component is compact and is
consistent with being the centre of activity (Taylor et al 1996).
CSOs often have a radio spectrum which peaks around a few GHz and thus
belong to the class of Gigahertz Peaked Spectrum (GPS)
sources. However this is not universally the case, for instance the
radio spectrum of the prototype CSO 2352+495 is quite flat
(e.g. Readhead et al 1996 (RPX96)). This turns out to be due to the
superposition of components which peak at different frequencies. GPS
radio galaxies all seem to be CSOs with relatively simple structures,
but more complex CSOs sometimes do not have a GPS spectrum. Most GPS
quasars do not seem to have CSO structures and may be a separate class
of object. The radio structures of the CSOs are similar to the large,
kiloparsec and megaparsec sized, double-sided radio sources (dubbed
'classical doubles') but are approximately 1000 times smaller.

The physical origin of Compact Symmetric Objects has been discussed
for many years. While soon after the detection of the first examples
it was suggested (Philips \& Mutel 1982) that they are young radio
sources which would evolve into large radio sources, alternative
suggestions have also been proposed: CSOs could be 'frustrated',
ie. located in a dense environment that could inhibit the growth of
the radio structure. CSOs could also be young radio sources, that will
'fizzle' out and die young (Readhead et al 1994) or stages of
intermittent radio activity (Reynolds \& Begelman 1997).

\section{Observations of Velocities in CSOs}
 
CSOs usually contain compact and bright compact `hotspot' components
located at the extremities of the source consistent with them being
the working surface of the jets as it propagates through the
ISM. Using multi-epoch VLBI observations it is possible to measure or
set limits on the rate of separation of their hotspots.

\subsection{Observational Summary}

The first upper limits on the rate of hotspot separation in CSOs
(Tzioumis et al 1989) showed that their velocities were
sub-relativistic and hence much smaller than the core-jet
objects. Conway et al (1994) measured possible sub-relativistic
motions in two CSOs but because these were based on only two epochs of
data they were not claimed as definite detections.  The first
unambiguous detections of CSO expansion were reported in the CSOs
0710+439 (Owsianik \& Conway 1998, OC98) and 0108+388 (Owsianik et al
1998, OCP98) based on multi-epoch VLBI observations over a decade or
more. Since then, detections or upper limits on expansion have been
determined for thirteen CSOs (Table \ref{cso_vel}). In the first two
parts of Table 1 we show the detections and limits on the
separation velocities of outer (hotspot) components.  For some sources
there are multiple speed estimates in the literature made at different
frequencies over different time intervals, in this case the one quoted
in Table 1 is the one with lowest error. The angular speed given is
the rate at which outer hotspot components increase their separation,
this is the relevant quantity for calculating sources ages.  A few
sources such as 1031+567 (Taylor et al 2000, TMP00) and 0108+388 (Owsianik et
al, in prep) may also have significant side-to-side motions, so the
total relative velocity between two hotspots quoted in some papers may
be larger than those in Table 1. Amongst the detections the rate of
expansion ranges between 0.1$h^{-1} \, c$ and $\sim$0.4$h^{-1} \, c$, the
unweighted mean value being 0.19$h^{-1} \, c$ (0.17$h^{-1} \, c$ 
including the limits).
% [{\bf if we include 4c31.04 then
% the highest velocity is 0.45 h-1c and the average is .21}]

In several cases motions have been detected for components which are
not at the edges of the source, these {\it internal} component
velocities are listed in the third part of Table 1.  In these cases we
may be measuring outward velocities of jet components.  These internal
velocities, as expected for a jet component origin, are larger than
for the hotspot components. In most cases these results are consistent
with the jet components moving with Lorentz $\gamma$ between 2-5, but
at relatively large angles to the line of sight.

   \begin{table}
      \caption[]{{\small Expansion Velocities and Kinematical ages of Compact Symmetric Objects}}
         \label{cso_vel}

\smallskip
         \begin{tabular}{llllrll}
%p{0.5\linewidth}  (h^{-1}c)
            \hline
            \noalign{\smallskip}
Source&z&Size$^{a}$&$v_{sep}^{a}$&Age(yrs)&Nr. of Epochs &Ref$^{b}$ \\
            \noalign{\smallskip}
            \hline
            \noalign{\smallskip}
\multicolumn{7}{c}{Detections} \\
            \noalign{\smallskip}

0035+227$^{c}$  &0.096&21.8&0.12$\pm$0.06&567&2 (1998-2001)& 1\\
0108+388 &0.669&22.7&0.18$\pm$0.01&417&5 (1982-2000)& 2 \\
%4C31.04&0.0592&70.1&0.448$\pm$0.0370&500&2 (1995-2000)&13 \\
0710+439 &0.518& 87.7 &0.30$\pm$0.02&932&7 (1980-2000)& 3 \\
1031+567$^{c}$  &0.4597&109.0&0.19$\pm$0.07&1836&2 (1995-1999)&4 \\
1245+676&0.1071&9.6&0.16$\pm$0.01&190&5 (1989-2001)& 5\\
OQ208      &0.0766&7.0&0.10$\pm$0.03&224&6 (1993-2002)& 6\\
1843+356$^{c}$  &0.763&22.6&0.40$\pm$0.04&180&2 (1993-1997)& 7\\
1943+546 &0.263&107.1&0.26$\pm$0.04&1306&4 (1993-2000)& 1 \\
2021+614  &0.227&16.1&0.14$\pm$0.02&368&3 (1982-1998)& 8 \\
2352+495  &0.238&117.3&0.12$\pm$0.03&3003&6 (1983-2000)& 9\\
            \noalign{\smallskip}
\multicolumn{7}{c}{Limits}\\
            \noalign{\smallskip}
1718-649&0.0142 &2.0& $<$0.07 &&2+&12  \\
1934-638&0.183&83.2& $<$0.05 &&3+& 10\\
1946+708&0.101&39.4& $<$0.10 &&5 (1992-1996)&11 \\
\multicolumn{7}{c}{Jet Components} \\
            \noalign{\smallskip}
            \noalign{\smallskip}
Source&z&Size$^{a}$&$v_{comp}^{a,d}$& ID$^{e}$ &Nr. of Epochs &Ref$^{b}$ \\
            \noalign{\smallskip}
\noalign{\smallskip}

0108+388 &0.669&22.70&0.7& C5&2 (1994-1997)& 4 \\
1031+567 &0.4597&109&0.6& &2 (1995-1999)&4 \\
1946+708&0.101&39.4&0.2& S2, S5&5 (1992-1996)&11 \\
1946+708&     &    &0.5-0.9& N2, N5&5 (1992-1996)&11 \\
2352+495  &0.238&117.3&0.4& B1a&6 (1983-2000)& 9\\
2352+495  &     &     &0.7& B5&2 (1994-1999)& 4\\
2352+495  &     &     &0.2& C1&6 (1983-2000)& 9\\

            \noalign{\smallskip}
            \hline
         \end{tabular}

{\small 

$^{a}$ The Linear size and the hotspot separation velocities
are reported in units of $h^{-1}$pc and $h^{-1}$c for H$_{o}=100h$ km
s$^{-1}$Mpc$^{-1}$
 
$^{b}$ References: 1. this paper, 2.  Owsianik et al (1998) and this
paper, 3. Owsianik \& Conway (1998) and this paper, 4. Taylor et al
(2000), 5. Marecki et al, this proc., 6. Stanghellini et al (2000),
7. Polatidis et al (2001), 8. Tschager et al (2000), 9. Owsianik et al
(2002), 10. Tzioumis et al (1989) and priv. com., 11. Taylor \&
Vermeulen (1997), 12. Tingay et al, these proceedings. 
  
$^{c}$  Velocity measurement between 2 epochs only, hence 
 provisional detection. 

$^{d}$ $v_{comp}$ Component velocity measured relative to the core, or
the source centre of symmetry.

$^{e}$ ID: Component name as identified in the relevant publication.
}
 \end{table}

\begin{figure}
\begin{center}
\centerline{\psfig{file=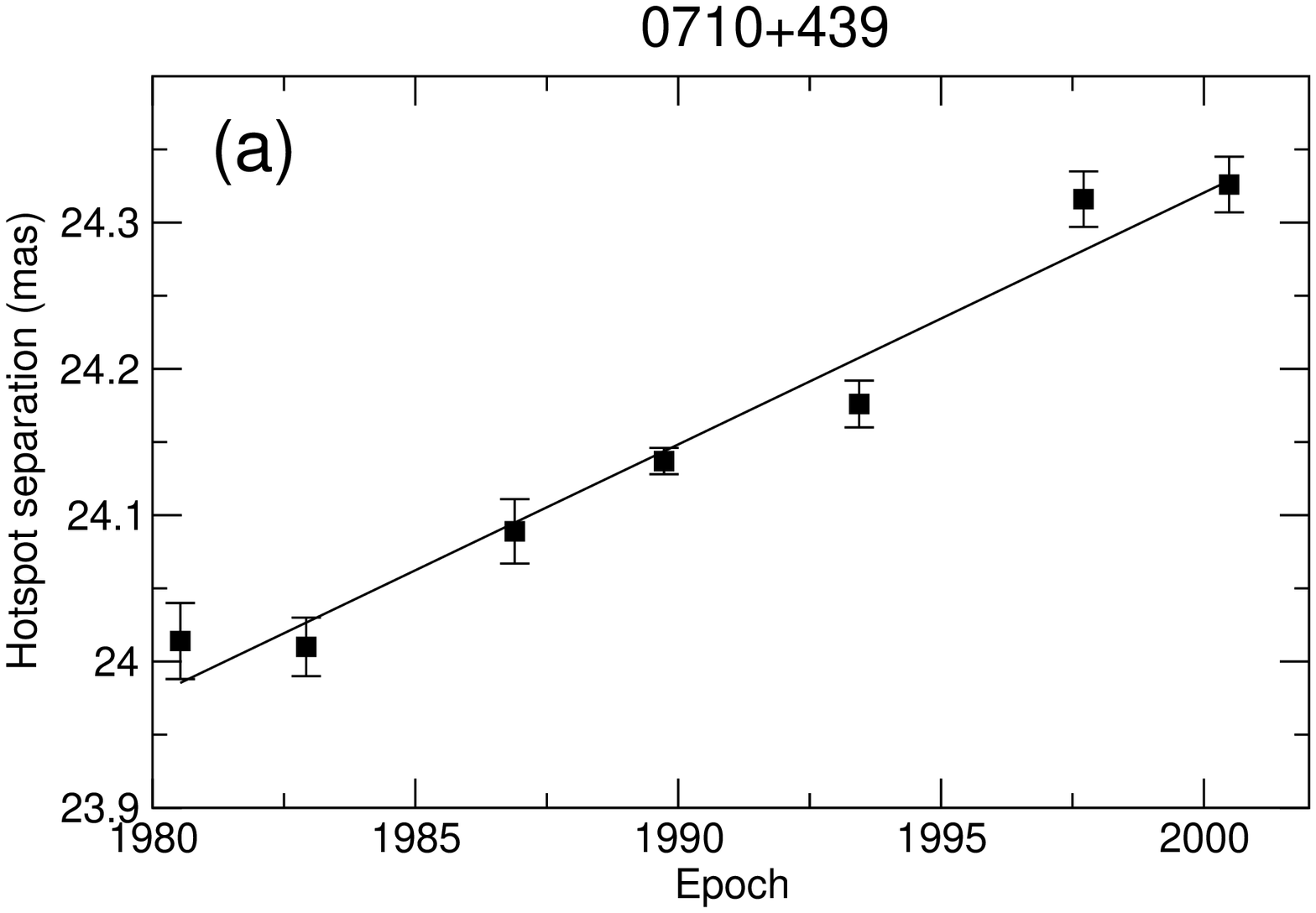,height=4.0cm}\psfig{file=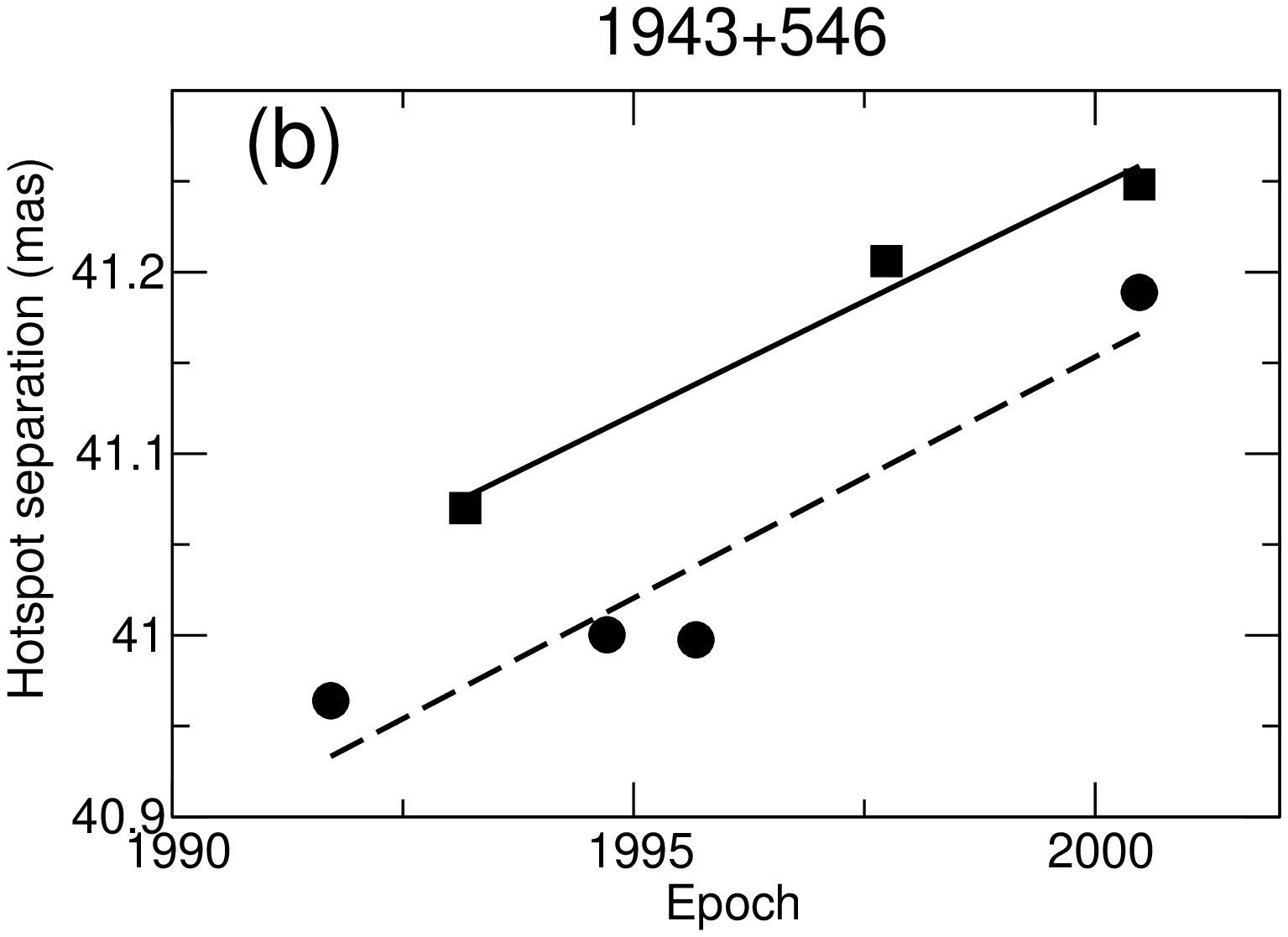,height=4.0cm}}

\caption{{\small (a) Hotspot separation versus time for CSO 0710+439
at 5 GHz. The best fit line represents an increase in the separation
of 17$\pm1 \mu$as/yr ($\rho$=0.983). Indicatory {\it a priori} error
bars representing the accuracy of the estimate of the separation are
plotted; they are derived from the beam size/SNR. (b) The hotspot
separation versus time for 1943+546 measured at 8 GHz (squares) and 5
GHz (circles). The best fit lines through the data have slopes of
25$\pm 2 \mu$as/yr (8GHz, $\rho$=0.987) and 27$\pm 2 \mu$as/yr (5GHz,
$\rho$=0.987).}}

\label{0710-1943vel}            % for cross-references
\end{center}
\end{figure}

The longest possible temporal coverage is obviously important in
getting good speed estimates (Table 1 shows the number of epochs and
the temporal coverage of the sources so far). In the sources 0710+439
(OC98), 0108+388 (OCP98) and 2352+495 (Polatidis et al 2002, POC02),
the $\lambda$6 cm VLBI observations cover by now almost 20 years and
consist of 5--7 measurements (epochs) per source (see Figure
\ref{0710-1943vel}a for CSO 0710+439).  With such data it is possible
to do meaningful regression analysis from which error bars on the
velocity can be estimated robustly. In these cases all the ordinary
least sqares methods  (e.g. Isobe et al 1990) gave consistent values
(withing the errors) for the rate of expansion. In contrast
measurements based on only a couple of epochs require error bars based
on {\it a priori} estimates of the accuracy to which component
positions can be measured (usually taken as 1/10 of a beam or the beam
size/signal-to-noise ratio). For long tracks, simple sources and high
enough map signal-to-noise on each component (i.e. SNR$ > 20$)
regression analysis for 0710+439 and 2352+495 shows that such {\it a
priori} position error estimates are plausible. However a minimum of
three epochs is probably required to feel fully confident of a real
velocity detection.

\subsection{Multiple Frequencies and Temporal Variations}

Measurements of the hotspot separation velocities in Table 1 have been
made at different frequencies; most often at 5 GHz (e.g. OC98, OCP98,
POC02, Tschager et al 2000), at 8.4 GHz (e.g Polatidis et al 1999 and
this paper) or 15 GHz (e.g. TMP00). Generally if a source has been
monitored at more than one frequency the derived velocities are
similar. For example, for 1943+546 we report an expansion velocity
(Fig \ref{0710-1943vel}b), from three epoch (1993.1-2000.4)
measurements at 8.4 GHz, of $v_{sep}=0.26\pm0.04 h^{-1}$c which is
within the errors with that independently derived using four epoch
(1991.7-2000.4) 5GHz VLBI observations ($v_{sep}=0.28\pm0.06
h^{-1}$c).

In the case of 2352+495, TMP00, based on two epoch 15GHz observations
(1994.9-1999.5), reported a velocity more than twice as large as the
20 year average at 5GHz (see POC02).  However restricting the analysis
to the three 5GHz epochs (and additional two 8GHz epochs) which cover
a similar time period as the two 15GHz epochs (1993.3-1997.7) a
consistent high velocity is found. The higher velocity found over a
short period may therefore be due to real temporal variations in the
hotspot advance speeds.

\subsection{Individual Hotspot Advance Speeds \label{differ}}

In cases where there is a strong, unambiguously identified core we can
attempt to measure the advance speed of each of the two hotspots
separately. In some cases apparent differences are seen.  For example
in 1943+546 while the eastern (and more distant from the core) hotspot
moves away from the core with $v_{hot}$=0.25$\pm 0.03 h^{-1}$c. the
nearer western hotspot is apparently barely moving (with a projected
velocity of 0.01$\pm 0.01 h^{-1}$c) away from the core.  Different
individual hotspot advance speeds are seen in a few other sources
where the core is identified (e.g. 0710+439, OC98).  These
measurements are very difficult, especially at 5GHz where CSO core
components are weak. We should remember that if weak jet components
are emerging from the core its apparent position can vary,
invalidating the separate core-hotspot velocities.

If different oppositely directed hotspot transverse velocities are
really being observed in 1943+546 and 0710+439 and assuming
simultaneous ejection from the core then it immediately implies that
hotspot advance speeds vary with time during the lifetime of the
source. If they were instead constant then the hotspot velocity ratio
would equal the hotspot-core-hotspot arm length ratio which is not the
case. For instance the two sigma lower limit on the hotspot velocity
ratio in 1943+546 is 6.8 yet the arm length ratio is only
1.68. Additional evidence for hotspot advance speed variability may
also be available from observations of 2352+495 (see Section 2.2).
Temporal variations in hotspot advance speed could be produced by
hydrodynamically introduced internal pressure changes or changes in
external density.  In hydrodynamic simulations through a smooth medium
Norman (1996) found variations in hotspot pressures, causing
variations in hotspot advance speed of about a factor of two. Such
different hotspot advance speeds are consistent with differences in
the hotspot pressures within CSOs of order 5 (e.g. RPX96,OC98). In
both 0710+439 and 1943+546 it is the highest pressure hotspot which is
moving fastest, as expected. However the magnitude of the difference
in advance speed in 1943+546 seems too large to explain by pressure
variations, in this case it is more likely due to variations in
external density. Perhaps in 1943+546 the eastern hotspot is moving
through an intercloud medium while the western hotspot is encountering
a cloud.

\subsection{Side-to-Side motions}

An assumption that is often made in source evolution models is that
the pressure of the hotspot is effectively distributed over a larger
area than that of the hotspot itself (the so-called ``dentist's
drill'' model, Scheuer 1982). In this model the hotspot has larger
side-to-side motions than its forward motion, and averaged over time
the area over which thrust is distributed is therefore
increased. Observations of CSOs seem to show that such side-to-side
motions are in fact much smaller than forward motions.(e.g POC02,
Polatidis et al in prep). The possible exceptions are 0108+388
(Owsianik et al in prep) and 1031+567 (TMP00). It is important in
future when reporting velocities to distinguish between the velocity
components along and perpendicular to the source axis. By detecting or
setting limits on perpendicular velocities the impact of side-to-side
motions on source evolution can be assessed.

\subsection{Expansion in Recurrent Sources}

In general, most CSOs have no large scale radio emission which might
be a sign of recurrent activity. There are however a couple of sources
in which weak extended emission has been detected, e.g. 0108+388
(e.g. Owsianik et al 1998), 1345+125 and OQ208 (Stanghellini et al
2002, these proceedings).  In addition to these cases there also exist
extreme examples of `double-double' sources in which the central
double has CSO-sized dimensions. For example the z=0.107 radio galaxy
1245+676 has a triple radio morphology (0.97 Mpc in extent, Lara et al
2001) and radio luminosity typical of an FR II galaxy except for
the fact that the central component dominates the flux density
($\sim$67\% at 1.4 GHz) and hence the total radio spectrum, and is
similar to the GPS sources. At parsec scale resolution, the core
appears as a Compact Symmetric Object. Its 9.6 parsec structure is
dominated by two mini-lobes containing hotspots; a slightly inverted
spectrum, weak component, located close to the centre of the structure
is tentatively identified with the core. VLBI observations at 5GHz
(1989.7-2001.5) have shown that the hotspots move apart with a
velocity of 0.163$\pm 0.008h^{-1}$c (Marecki et al, these
proceedings). This implies a kinematic age of 190 years for the core
region. 1245+676 is by far the best example where the CSO appears to
be the youngest phase of recurrent radio activity, hinting that at
least some CSOs may be re-born radio sources.

\section{Discussion}

In the previous section we discussed the observations and specific
sources. In this section we discuss more generally how the proper
motions constrain the general properties of the CSO population.

\subsection{Kinematic Ages of CSOs}

The most direct result of the CSO expansion measurements are the low
kinematic ages derived by dividing the projected source size by the
measured projected separation velocities (see Table 1) which are all
$\leq 3 \times 10^{3}$ yrs. What is important here is not the exact
number (which may be revised as new measurements are added) but rather
the order of magnitude.  This implies that CSOs are {\it young}
objects. This constitutes so far the most direct way to estimate the
age of an extragalactic radio source.

Before accepting these age estimates we should consider whether the
measured source expansions (which represent the {\it instantaneous}
hotspot separation rate) are truly representative of the {\it mean
growth rate} of the sources.  CSOs might conceivably expand in brief
bursts when encountering a relatively low density medium, their
advance being hindered by jet-cloud interactions during the rest of
the time.  If this were happening we would be measuring only the
velocity of these brief expansion periods and hence severely
underestimating their age.  This is highly unlikely, given that we
have measured expansion speeds in a very high fraction of the sources
where we have good data (10 of 13 cases, Table 1). This implies
that the instantaneously measured separation velocities are consistent
with the {\it mean hotspot separation speed} and that the kinematically
estimated ages are an accurate representation of the radio source
lifetime.

\subsection{Other Age Estimates and Equipartition \label{age_est}}

Age estimates for CSOs have also been made by indirect means.
Readhead et al (1996) applied the classical 'waste energy basket'
argument to 2352+495 and derived an age of $\sim$3000 years, which is
similar to the kinematical age of 3003 years (POC02). There have also
been attempts to estimate the age of CSOs and the larger double
sources (MSOs) via the detection of high frequency breaks in their
spectra due to ageing of the electrons in the lobes. Minimum energy
and equipartition conditions are also usually assumed.  The estimated
spectral ages (eg. RPX96, Murgia et al 1998) are 10$^{3}-10^{4}$
years, similar to the kinematic ages. In fact Murgia (these
proceedings) derives a spectral age for 1943+546, very close to the
kinematic age of 1297 years.

The close agreement of kinematic and other age estimates suggests
that most CSOs are indeed young radio loud sources. The agreement also shows
that particle and fields are probably close to equipartition in CSOs and 
that the standard model of radiative ageing is roughly correct.
These are important results since it is very unclear whether
equipartition and standard spectral ageing apply in classical double
sources (see Blundell and Rawlings 2000, Rudnick 2002).

\subsection{Velocity Correlations }

\begin{figure}
\begin{center}
\centerline{\psfig{file=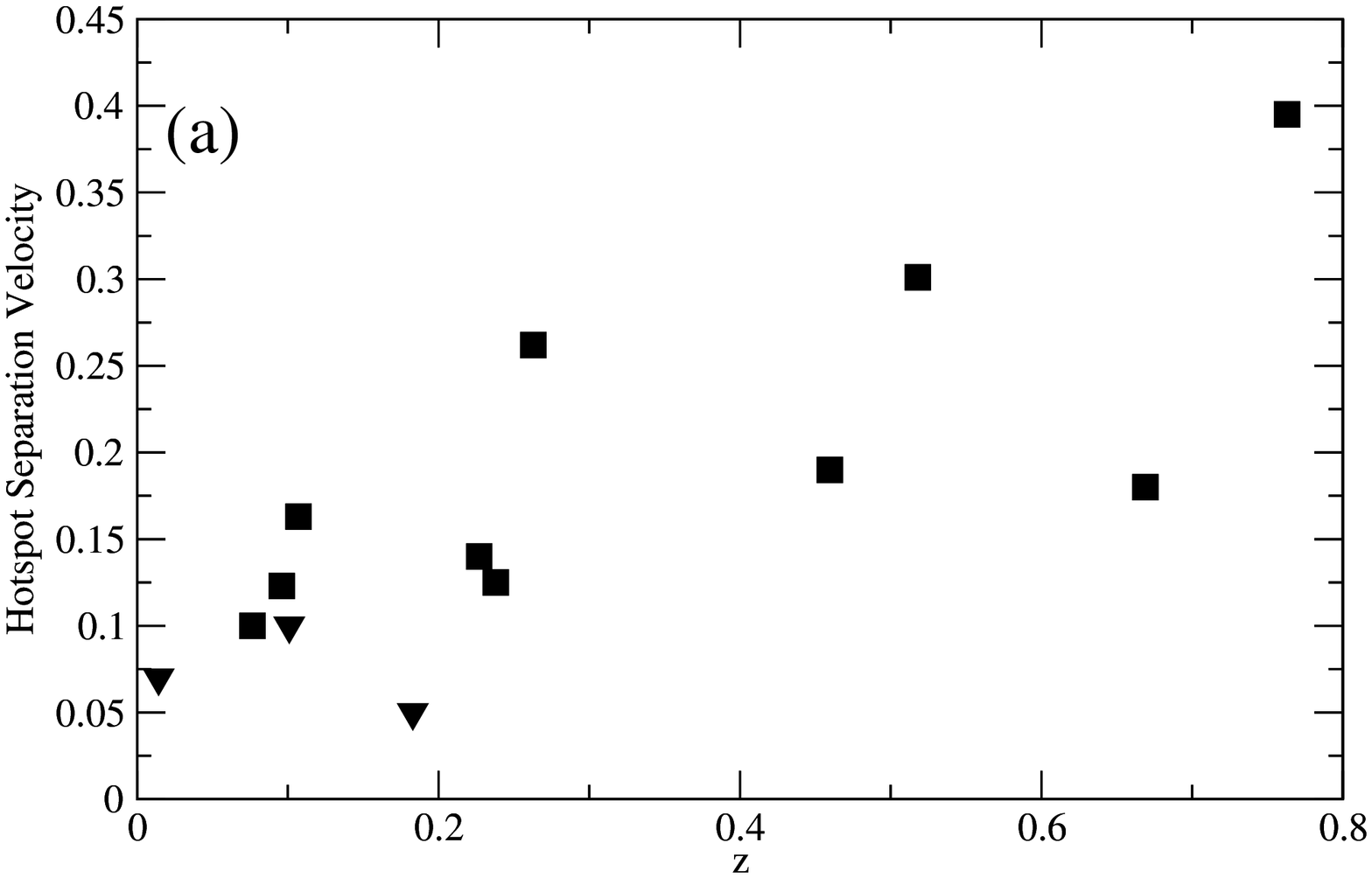,height=3.8cm}\psfig{file=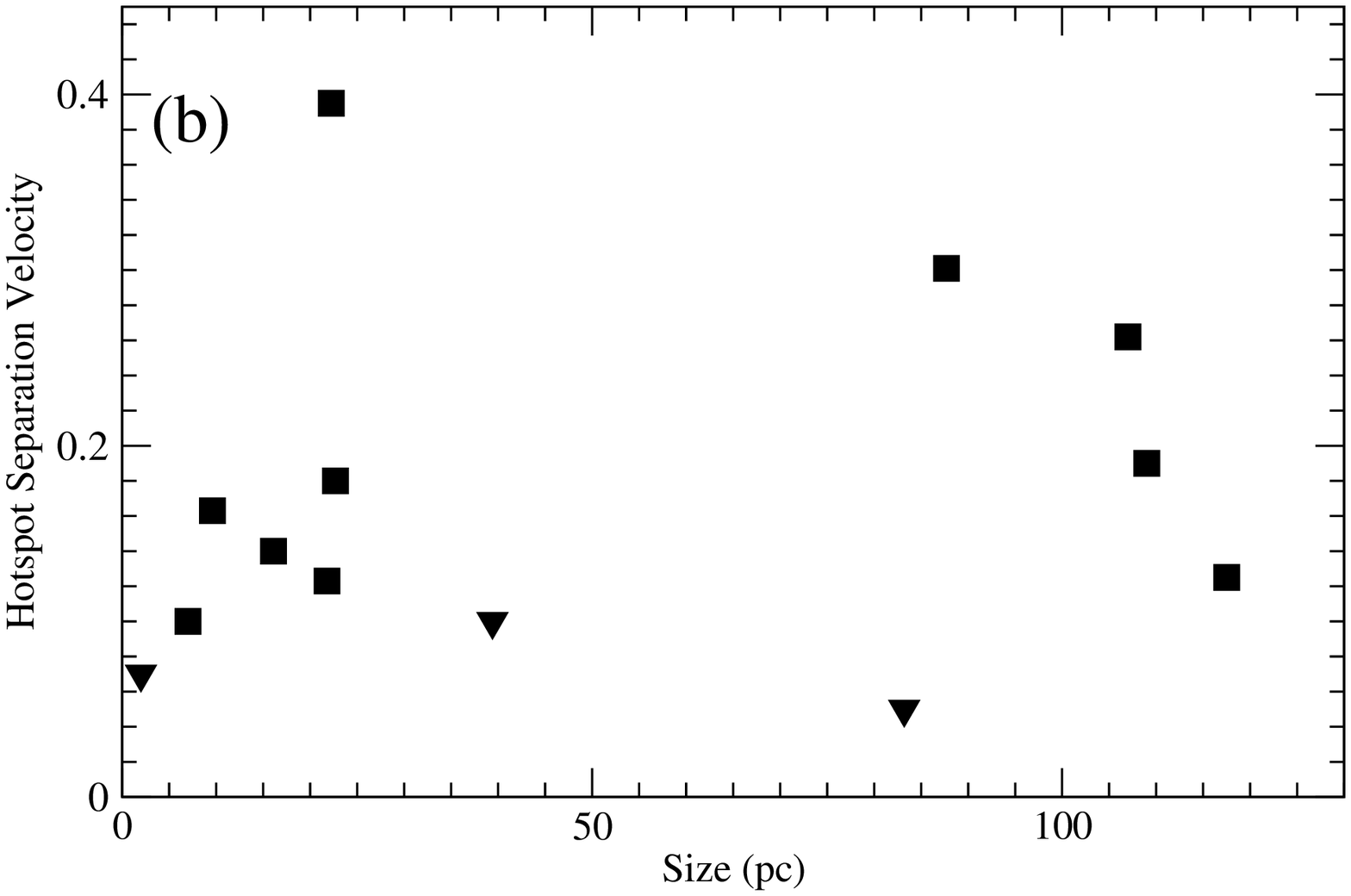,height=3.8cm}}

\caption{{\small Distributions of (a) Hotspot expansion velocity
vs. redshift (b) The hotspot expansion velocity vs. projected linear
size (squares are the detections, triangles are the upper limits).}}
\label{correlfig}
\end{center}
\end{figure}

We have searched for possible correlations (Polatidis et al, in prep)
between separation velocity and luminosity, redshift, source size and
arm length ratio (see Sec \ref{arm-length}). We find an apparent
correlation with redshift (with a correlation coefficient
$\rho$=0.752) but no correlation with source size (see Fig
\ref{correlfig}) or luminosity. However this apparent correlation
might be observationally biased. Slowly expanding sources will be hard
to detect at larger z, so we would not expect the bottom right of the
velocity-z graph to be filled.

\subsection{Hotspot Advance Speeds and Source Orientations \label{arm-length}}

Proper motion observations obviously give only the velocities
projected on the sky plane. We would like to estimate the hotspot
advance speeds through their surrounding medium which can in turn be
used to constrain external densities via ram pressure arguments.  The
absence of superluminal motions and relativistic beaming together with
the edge brightened source morphologies seems consistent with CSOs
being isotropically orientated. In this case the projected hotpot
advance speeds are on average half of their total speeds. Hence the
mean advance speed of {\it each hotspot} through its external medium
would equal the mean projected hotspot separation velocity or 0.19$h^{-1}
\, c$ (see Sect 2.1).

\begin{figure}
\begin{center}
\centerline{\psfig{file=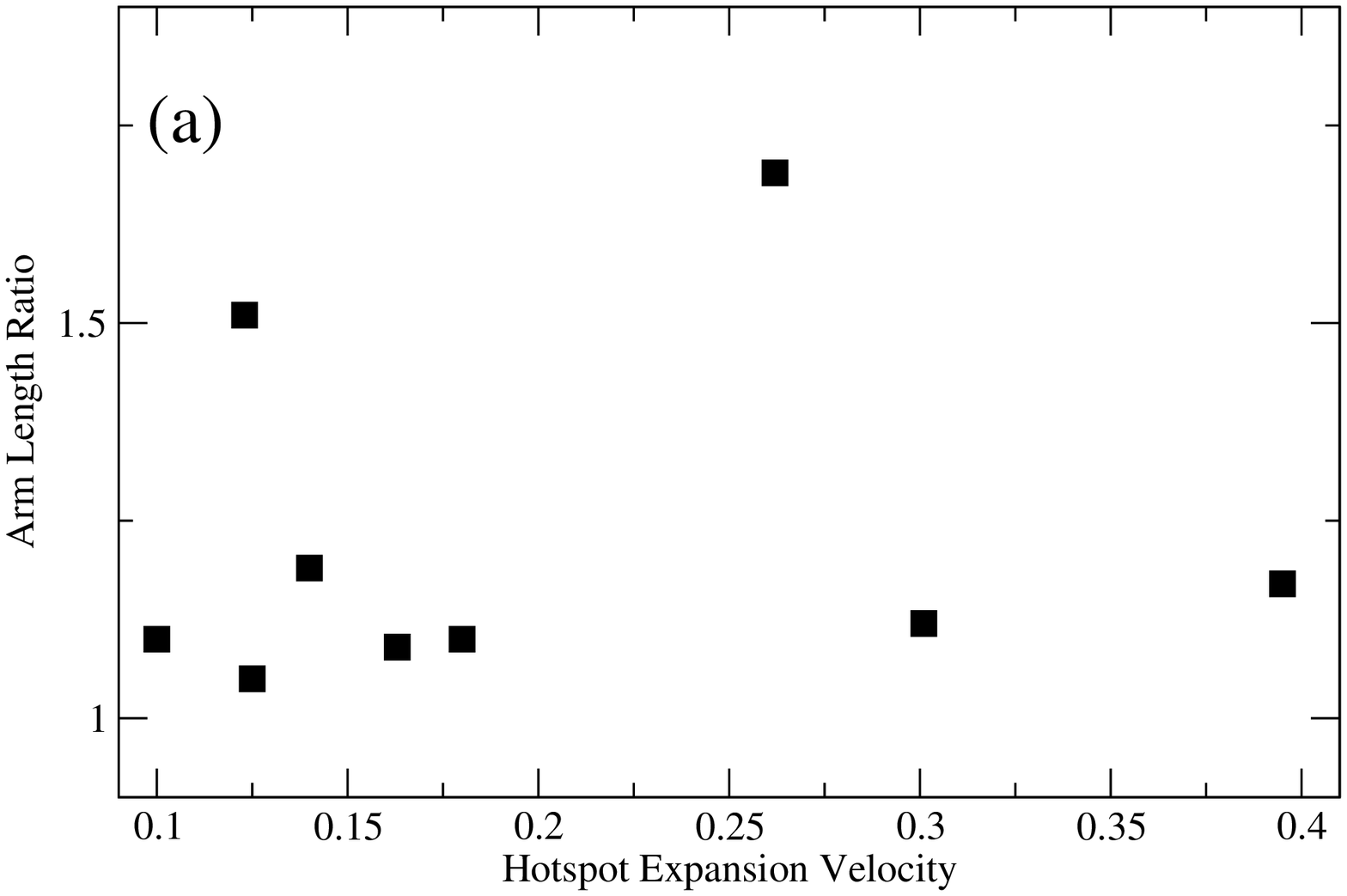,height=3.8cm}\psfig{file=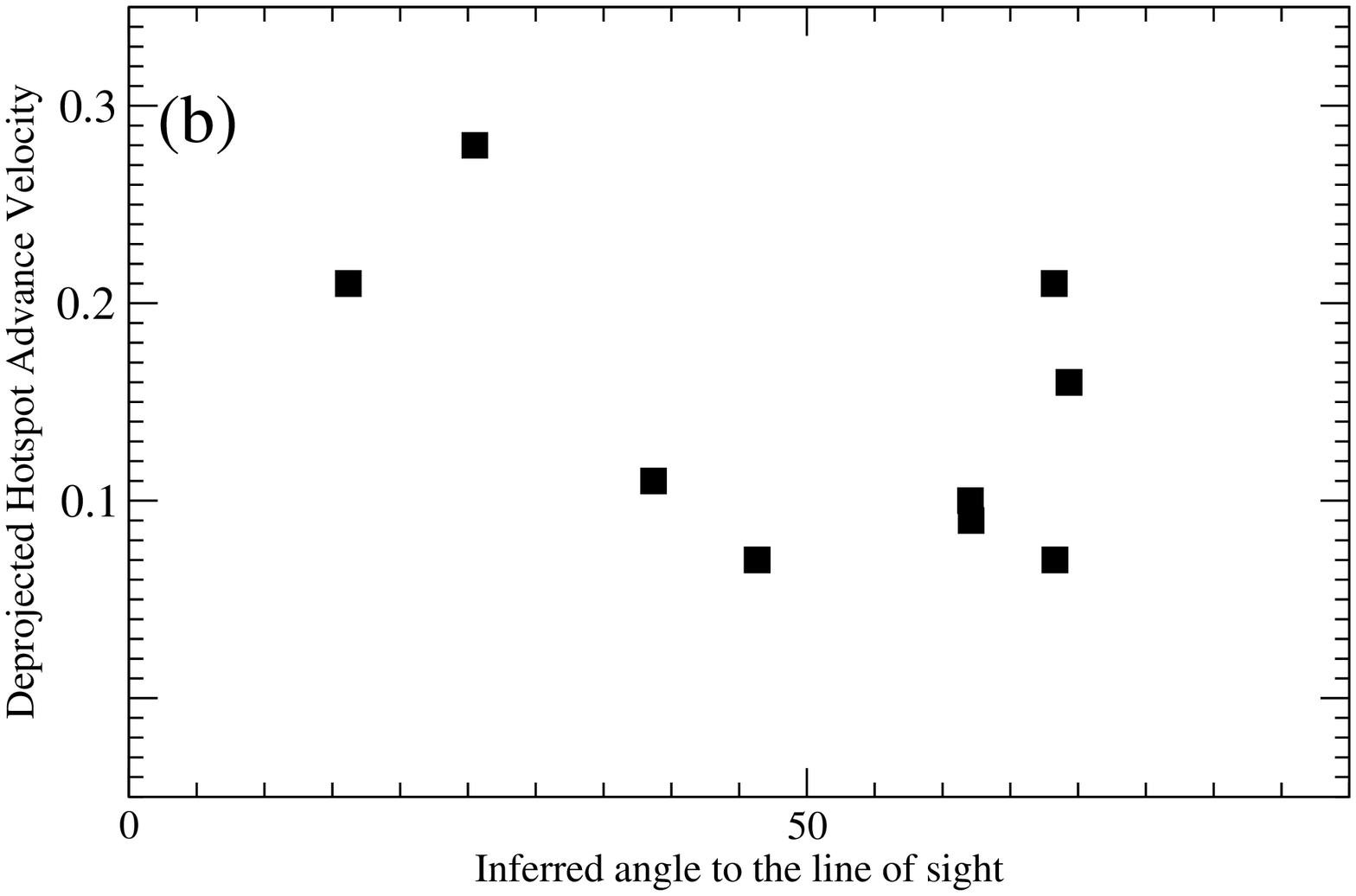,height=3.8cm}}
\caption{{\small (a)The arm-length ratio vs. the hotspot expansion velocity (b) The derived deprojected hotspot advance velocity vs. the inferred angle to the line of sight of the radio source axis (see Section \ref{arm-length}). }}

\label{armfig}            % for cross-references

\end{center}
\end{figure}

Assuming only light travel time effects, one could try to derive the
deprojected hotspot advance velocity and the angle to the line of
sight using both the arm-length ratio and the observed velocities. Fig
\ref{armfig}a shows the measured arm-length rations and hotspot
separation velocities for the nine sources with a detection in Table 1
for which the core has been identified (except for 1031+567). Figure
\ref{armfig}b shows the deprojected hotspot advance velocities and the
inferred angle to line of sight assuming all the asymmetry is due to
light travel time effects.  Of course it is likely that part of the
arm-length asymmetry is intrinsic and not light travel time induced.

\subsection{Evolution of CSOs}
 
 Measured CSOs expansions show that they are young sources, however
the subsequent evolution of these sources is less clear.  The simplest
assumption is that CSOs evolve into classical double sources like
Cygnus A. Alternatively CSOs could comprise a population of short
lived sources which 'fizzle-out' after a short lifetime. Answering the
question of CSOs subsequent evolution requires studying the population
densities of different sizes of source. However the CSO velocities
also provide some constraint on models, because subsequent evolution
cannot give significantly larger hotspot velocities or else
relativistic effects would be seen. In addition the velocity
measurements, assuming ram pressure confinement of the hotspots by the
ISM, constrain the external densities at distances of a few to a few
hundred parsecs from the centre of activity to be of order 1 cm$^{-3}$
(e.g. RPX96, OC98, Conway 2002) which implies that the external
density is not power law down to parsec scales but rather has a King
profile with scale length of order 1kpc. Such a turnover may also
explain the redshift distribution of GPS and larger sources (Snellen
et al 1999).

O'Dea \& Baum (1997)  found that the relative number of 
MSOs and large scale sources were roughly consistent 
with an evolution model in which sources expand into 
a medium with decreasing density with radius and  undergo
the expected negative luminosity evolution. However in this model 
given the external density  turnover at $<1$kpc one would expect 
there  to be very  much fewer sub 100pc sized sources than are 
observed. Such an excess may imply that CSOs belong mainly 
to a separate population of short lived sources, however 
it is possible that luminosity selection  effects influence this 
result (viz  Snellen et al 1999). 
An alternative explanation  is that for some reason small CSOs are 
closer  to equipartition (see  Section \ref{age_est}) than larger sources, 
which is an additional effect increasing  their  efficiency of 
converting  jet energy to radio luminosity, which then  boosts 
their representation  in flux limited samples (Conway 2002).

\section{Conclusions}

The combination of proper motion studies in CSOs with other 
lines of evidence strongly argues  that most CSOs
are young objects. One of the main questions posed at the 
last GPS conference therefore seems to be answered. 

A fundamental consequence of such CSO youth,    
which is often overlooked, is that narrow jets and hotspots 
apparently exist  only a few hundred years after the 
start of the radio-loud activity. This very short characteristic time 
must  strongly constrain the scales and mechanisms by which jets
are formed and accelerated. If the source activity started with a
wide angle wind or slowly accelerating outflow we would not see 
the CSO morphologies that we observe.  One can summarize this
as  'Jet activity starts like  an electric motor and not like a 
steam engine' (R.Blandford, private communication).

Having established the basic fact that CSOs are young, proper motion
studies still have a lot to contribute. For instance its not yet clear
whether {\it all} CSOs are young or if some are frustrated. Continued
monitoring of sources with upper limits is therefore
important. Finding the distribution of CSO velocities and looking for
correlations with other quantities is another important goal to
constrain evolution models. Long term monitoring can also reveal or
set limit on hotspot accelerations/decelerations, different hotspot
advance speeds in the same source or hotspot side-to-side motions; all
are useful to constrain the dynamics of how radio sources evolve.

The rate at which new CSO motions are reported in the literature 
is  encouraging for answering the above questions. 
We should however remember to continue to be very careful
in our measurements and our interpretations (and remain very 
patient!) since detecting these very small angular velocities 
remains technically challenging.

\section*{References}

\reference Blundell K.M., Rawlings, S., 2000, AJ, 119, 1111
 
\reference Conway, J.E., et al 1992, ApJ,  396, 62.
%Pearson, T.J., Readhead, A. C.S., Unwin, S.C., Xu, W.,Mutel, R.M.,
 
\reference Conway, J.E., et al, 1994, ApJ, 425, 568.
% Myers, S.T., Pearson,
%T.J., Readhead, A. C.S., Unwin, S.C. , Xu, W.
 
\reference Conway, J.E., 2002, NewAR, 46, 2-7,  263

\reference Isobe, T., et al, 1990, ApJ, 364, 104 

\reference Lara, L., et al, 2001, A\&A, 370, 409
%Cotton, W. D., Feretti, L., Giovannini, G., Marcaide, J. M., Márquez, I., Ventu 

\reference Marecki A., 2002, these proceedings

\reference Murgia M., et al 1999, A\&A, 345, 769.
%Fanti, C., Fanti, R.,1999, Gregorini L.,Klein U., Mack K.-H., Vigotti M.,
 
\reference Murgia M., 2002, these proceedings

\reference Norman M., 1996, in 'Energy Transport in radio galaxies and Quasars'
PASP Conf Vol 100. p 405.

\reference O'Dea C.P., 1998, PASP, 110, 493.
 
\reference Owsianik, I., Conway, J, 1998, A\&A,  337, 69 (OC98).
 
\reference Owsianik, I., Conway, J., Polatidis, A., 1998,
A\&A,  336, L37. (OCP98)

\reference  Polatidis, A.G., Owsianik, I., Conway, J.E., 2002,
A\&A, sub. (POC02)

\reference Phillips, R.B., Mutel, R.L., 1982, A\&A,  106, 21

\reference Polatidis, A.G.,  et al 1999, NewAR, 43/8-10, 657
 %Wilkinson P.N., Xu. W., Readhead A.C.S., Pearson T..J., Taylor G.B., Vermeulen 

\reference Polatidis, A.G., 2001, in "Similarities and Universalities of
Relativistic Flows", eds. M Georganopoulos et al., Logos Verlag, 96

\reference Readhead A.C.S., et al 1994, in Proc.NRAO Workshop 23, 17
% Xu W., Pearson T.J., Wilkinson P.N., Polatidis A.G.,
%1994, in Proc. NRAO Workshop 23:Compact
%Extragalactic Radio Sources, ed. J.A. Zensus and K.I. Kellermann,
%(Socoro: NRAO), 17.

\reference Readhead A.C.S., et al , 1996, ApJ, 460, 612. (RPX96)
% Taylor G.B., Xu W., Pearson T.J.,Wilkinson P.N., Polatidis A.G.

\reference Reynolds, C.S., Begelman M.C., 1997, ApJ, 487, L135.

\reference Rudnick  L., 2002, NewAR , 46, 2-7, 95.

\reference Scheuer P., 1982, In Proc. IAU Symp 97, 163

\reference Stanghellini C., Liu, X.,Dallacasa, D., Bondi, M.,  2002a, 
NewAR, Vol 46, 287.

\reference Stanghellini C., et al 2002b, these proceedings

\reference Taylor, G.B., et al, 1996, ApJ, 463, 95 
%Readhead, A.C.S., Pearson, T.J., 1996, ApJ, 463, 9

\reference Taylor, G.B., Vermuelen, R.C., 1997, ApJ, 485, L9

\reference Taylor, G.B., et al, 2000, ApJ, 541, 112 (TMP00)

\reference Tschager, W. et al, 2000, A\&A, 360, 887.
%Schilizzi, R.T., Rottgering, H. J. A., Snellen, I.A.G., Miley, G.K.,

\reference Tzioumis, A.K., et al, 1989, AJ, 98, 36

\reference Wilkinson, P.N., Polatidis, A.G., et al 1994, ApJ,  432, L87.
%Readhead, A.C.S., Xu, W., Pearson, T.J.,

\end{document}